\documentclass[osajnl,twocolumn,showpacs]{revtex4}
\usepackage{graphicx}

\begin{document}

\title{Condensation of classical nonlinear waves}

\author{Colm Connaughton$^1$, Christophe Josserand$^2$, Antonio Picozzi$^3$, Yves Pomeau$^1$ and 
Sergio Rica$^{4,1}$}
\affiliation{${}^1$Laboratoire de Physique Statistique, ENS-CNRS, 24 rue Lhomond,
F75005 Paris-France\\
$^2$Laboratoire de Mod\'elisation en M\'ecanique,
CNRS UMR 7607, Case 162, 4 place Jussieu, 75005 Paris-France\\
$^3$LPMC-CNRS, Universit\'e de Nice Sophia-Antipolis, Nice, France.\\
$^4$Departamento de F\'\i sica, 
Universidad de Chile,
Blanco Encalada 2008, 
Santiago, Chile.}

\begin{abstract}
We study the formation of a large-scale coherent structure (a condensate) in classical wave equations by considering the defocusing nonlinear Schr\"odinger equation as a representative model. We formulate a thermodynamic description of the condensation process by using a wave turbulence theory with ultraviolet cut-off. In 3 dimensions the equilibrium state undergoes a phase transition for sufficiently low energy density, while no transition occurs in 2 dimensions, in analogy with standard Bose-Einstein condensation in quantum systems. 
Numerical simulations show that the thermodynamic limit is reached for systems with $16^3$ computational modes and greater. 
On the basis of a modified wave turbulence theory, we show that the nonlinear interaction makes the transition to condensation subcritical. The theory is in quantitative agreement with the simulations. 

\end{abstract}
\pacs{42.65.Sf,05.45.-a,47.27.-i}
\maketitle

The problem of self-organization in conservative systems has generated 
much interest in recent years. For infinite dimensional Hamiltonian systems like
{\it classical} wave fields, the relationship between formal reversibility and
actual dynamics can be rather complex.
Nevertheless, an important insight was obtained from {\it soliton} dynamics, in particular through numerical 
simulations performed for the focusing nonintegrable nonlinear Schr\"{o}dinger (NLS) equation 
\cite{Pet-85,Zakharov-88}.
These studies revealed that this system would generally evolve to a state 
containing a large-scale coherent localized structure, 
{\it i.e.}, solitary-wave, immersed in a sea of small-scale turbulent fluctuations.
The solitary-wave plays the role of a ``statistical attractor" for the 
system, while the small-scale fluctuations contain, in principle, 
all information necessary for time reversal.
Importantly, the solitary-wave solution corresponds to the solution that 
minimizes the energy (Hamiltonian), so that the system tends to relax towards 
the state of lowest energy, while the small-scale fluctuations compensate for 
the difference between the total energy and that of the coherent structure~\cite{Zakharov-88}.

It is only recently that a statistical description of this self-organization process has been elaborated~\cite{Jordan-00,Rumpf-01}.
Remarkably it turns out that when the Hamiltonian system is constrained by an additional invariant
({\it e.g.}, the mass), the increase of entropy of 
small-scale turbulent fluctuations {\it requires} the formation of coherent structures to ``store" the invariant~\cite{Rumpf-01}, 
so that it is thermodynamically advantageous for the system to approach the ground state which minimizes the energy~\cite{Zakharov-88}.

In the present work we consider the {\it defocusing} regime of the NLS dynamics, 
which is also relevant to the description of thermal Bose gases~\cite{Davis-01,Davis-prl,Berloff-02}.
This regime would be characterized by an irreversible evolution of the system towards a homogenous plane-wave~\cite{Dyachenko-92,Pomeau-92}, as described by the weak turbulence theory~\cite{ZakhBook}.
This evolution is consistent with the general rule discussed 
above~\cite{Zakharov-88,Jordan-00,Rumpf-01}, because the 
plane-wave solution realizes the minimum of the energy (Hamiltonian) in the defocusing case. 
Accordingly, the NLS equation would exhibit a kind of 
condensation process, a feature that has been accurately confirmed only
recently in the context of thermal Bose fields, where intensive numerical 
simulations of the 
NLS equation have been performed in 3$D$~\cite{Davis-prl,Berloff-02}.

In this Letter we formulate a thermodynamic description of this condensation process. 
We show that the 3D-NLS equation exhibits a {\it subcritical} condensation 
process whose thermodynamic properties are analogous to those of Bose-Einstein 
condensation in quantum systems, despite the fact that the considered 
wave system is completely {\it classical}.  Our analysis is based on the 
kinetic theory of the NLS 
equation, in which we introduce a frequency cut-off to regularize the 
ultraviolet catastrophe inherent to ensembles of classical 
waves (Rayleigh-Jeans paradox).
This allows us to clarify the essential role played by the number of modes in 
the dynamical formation of the condensate.
Furthermore, we find that in 2$D$ the NLS equation does not undergo 
condensation in the thermodynamic limit,
in complete analogy with a uniform, ideal Bose gas.
The significance of this result is that the system irreversibly evolves towards a state of equilibrium (maximum entropy) {\it without generating a coherent structure that minimizes the energy} (Hamiltonian), in contrast with the general rule commonly accepted~\cite{Zakharov-88,Pomeau-92,Jordan-00,Rumpf-01}.

Given the universal character of the NLS equation in nonlinear wave 
systems, the reported condensation process could stimulate interesting new 
experiments in various branches of physics. Nonlinear optics is a natural 
context where classical wave condensation may be observed and studied 
experimentally, in relation to the recent proposal of condensation for 
photons~\cite{Chiao}. Moreover, the formal reversibility of the condensation 
process could be demonstrated by means of an optical phase-conjugation 
experiment~\cite{Boyd}.
In addition to optics, wave condensation could be relevant for hydrodynamic 
surface waves~\cite{zakhprl04},
on the basis of the  recent progress on measurements of Zakharov's spectra. 
Moreover, our work is also relevant to the important issue of strong wave 
turbulence~\cite{Robinson-97}, where the emergence and persistence of large 
scale coherent structures in the midst of small scale fluctuations is a common 
feature of many turbulent fluids, plasmas and optical wave systems.

We consider the normalized defocusing NLS equation in $D$ spatial dimensions 
for the complex function 
$\psi$\cite{GP}:
\begin{equation}
i \partial_t \psi \, = \, -\Delta \psi \, + \, |\psi|^2 \psi
\label{NLS}
\end{equation}
\noindent
where $\Delta$ stands for the Laplace operator in dimension $D$. 
This equation describes the evolution of defocusing interacting waves through the cubic nonlinear term. 
The dynamics conserves the mass (particle number) $N = \int  |\psi|^2  d^Dx $, and the 
total energy $H = \int \left( |\nabla\psi |^2   +  \frac{1}{2}  |\psi|^{4 }\right) d^Dx $.

We address the dynamical formation of the condensate starting from a non-equilibrium stochastic initial condition for the wave $\psi$, which we take to be of zero mean and statistically homogenous. 
In spite of the formal reversibility of the NLS Eq.(\ref{NLS}), the nonlinear 
wave $\psi$ may exhibit an {\it irreversible} evolution towards thermal 
equilibrium. The salient properties of this evolution may be described by 
weak-turbulence theory. This theory derives, within certain
approximations~\cite{ZakhBook}, an irreversible 
kinetic  
equation for the averaged wave-spectrum,  
$\left< a_{{\bf k}_1} a_{{\bf k}_2}^* \right> = n_{{\bf k}_1} \delta^{(D)}({\bf k}_1-{\bf k}_2)$, $a_{\bf{k}}$ being 
the Fourier transform of $\psi$ defined by $a_{\bf k} (t) =  \int  \psi({\bf x},t) \, e^{-i {\bf k}
\cdot {\bf x}} d^Dx $
\begin{eqnarray}
\nonumber \partial_t n_{{\bf k}_1}(t) =  {\cal C}oll[n_{{\bf k}_1}]\equiv \int \!\!d^Dk_2 d^Dk_3 d^Dk_4 W_{k_1,k_2;k_3,k_4}  \ \ \ \ \ \ \ \ \ \ \  \ \ \ \ \ \ \ \\
 \ \ \ \ \ \ \ \ \ \ \ \ \ \ \ \ \ \ \ \times \ \left(n_{{\bf k}_3}n_{{\bf k}_4}n_{{\bf k}_1}\!+\! n_{{\bf k}_3}n_{{\bf k}_4} n_{{\bf k}_2}\!-\!
n_{{\bf k}_1}n_{{\bf k}_2}n_{{\bf k}_3}\!-\!n_{{\bf k}_1}n_{{\bf k}_2} n_{{\bf k}_4}\right) 
\label{boltz}
\end{eqnarray}
with $W_{k_1,k_2;k_3,k_4} = \frac{4\pi}{(2\pi)^{D}} 
\delta^{(D)}({\bf k}_1+{\bf k}_2-{\bf k}_3-{\bf k}_4)$ $ 
\delta^{(1)}(k_1^2+ k_2^2-k_3^2-k_4^2)$~\cite{ZakhBook}. 
As for the usual Boltzmann's equation, Eq.(\ref{boltz}) conserves the 
mass $N=V\int n_{{\bf k}}(t) d^Dk $,  
the kinetic energy $E=V\int k^2 n_{{\bf k}}(t) d^Dk $ (being $V$ the system volume for $D=3$ and the surface for $D=2$), and exhibits a $H$-theorem of entropy growth $d \mathcal S/dt \ge 0$, where $ \mathcal S(t)=\int  \ln(n_{{\bf k}}) \, d^Dk$ is the non-equilibrium entropy.
Accordingly, the kinetic equation (\ref{boltz}) describes an irreversible evolution 
of the wave-spectrum 
towards the Rayleigh-Jeans {\it equilibrium} distribution~\cite{ZakhBook}: 
\begin{equation}
n^{eq}_k= \frac{T}{k^2 -\mu},
\label{equil}
\end{equation}
where $T$ and $\mu(\le 0)$ are, by analogy with thermodynamics, the 
temperature and the chemical potential respectively.
The spectrum (\ref{equil}) is lorentzian and the correlation length of the 
wave $\psi$ is determined by $\mu$, 
$\lambda_c \simeq 1/\sqrt{-\mu}$.

The distribution (\ref{equil}) realizes the maximum of the entropy 
${\mathcal S}[n_{\bf k}]$ and vanishes exactly the collision term, ${\cal C}oll[n^{eq}_{k}]=0$.
However, it is important to note that Eq.(\ref{equil}) is only a {\it formal} 
solution, because it does not lead to converging expressions for the energy 
$E$ and the mass $N$ in the short-wavelength limit $k \rightarrow \infty$.
To regularize this unphysical divergence, we introduce an ultraviolet cut-off 
$k_c$, i.e., we assume $n_{{\bf k}}(t) = 0$ for $k > k_c$. Note that 
this cut-off arises naturally in numerical simulations through the 
spatial discretization of the NLS equation, and manifests itself in real 
physical systems through viscosity or diffusion effects at the microscopic 
scale \cite{ZakhBook}.

To begin let us analyze the equilibrium distribution (\ref{equil}) in 3$D$.
From energy and mass conservation one gets 
\begin{eqnarray}
\frac{N}{V} = 4 \pi \, T k_c \left[1 - \frac{ \sqrt{-\mu}}{k_c} \arctan\left( \frac{k_c}{\sqrt{-\mu}} \right) \right]    \label{nn}\\
\frac{E}{V}  =    \frac{4 \pi \, T k_c^3 }{3} \left[ 1+ 3 \frac{ \mu}{k_c^2}   + 3 \left(\frac{ -\mu}{k_c^2} \right)^\frac{3}{2}\arctan\left(\frac{k_c}{\sqrt{-\mu} }\right) \right]   \label{ee}
\end{eqnarray}
These equations should be interpreted as follows. 
The initial non-equilibrium state of the field $\psi$ is characterized by a 
mass $N$ and an energy $E$. 
The wave spectrum then relaxes to
the equilibrium distribution (\ref{equil}), whose temperature $T$ and chemical 
potential $\mu$ are determined by Eqs.(\ref{nn},\ref{ee}): a given pair $(N,E)$
then determines a unique pair $(T,\mu)$.
An inspection of Eq.(\ref{nn}) reveals that $\mu$ tends to $0$, i.e. $\lambda_c$ diverges to infinity, for a non-vanishing temperature $T$ (keeping $N/V$ constant), or a finite density $N/V$ (keeping $T$ constant).
By analogy with the Bose-Einstein transition in quantum systems, 
this reveals the existence of a condensation process. 
The same conclusion follows from the analysis of the energy per particle $E/N$.
There exists a non-vanishing critical energy per particle $E_{\rm tr}/N  = k_c^2/3$
such that $\mu=0$.
Dividing (\ref{ee})  by (\ref{nn}), one gets $(E- E_{\rm tr})/(N k_c^2) = \pi 
\sqrt{-\mu}/(6 k_c) +{\cal O} \left(-\mu/k_c^2\right)$. Decreasing the energy 
per particle effectively cools the system, and one reaches a finite 
threshold $E_{\rm tr}$, below which condensation occurs. 

The same analysis in 2$D$ readily gives $N/V=\pi T \ln(1-k_c^2/\mu)$ and $E/N = \mu+k_c^2/\ln(1-k_c^2/\mu)$. In this case 
$\mu$ reaches $0$~: $(i)$ for a vanishing temperature $T$ ($N/V$ constant), $(ii)$ for a diverging density $N/V$ ($T$ constant), $(iii)$ for a vanishing energy per particle $E/N$. It results that condensation no longer take place in $2D$. We have confirmed this result by direct numerical simulations of the wave equation (\ref{NLS}).
In contrast to the 3$D$ case where the field $\psi$ generates a coherent plane-wave (condensate), in 2$D$ $\psi$ remains a stochastic field of zero mean, whose spectrum evolves towards the equilibrium distribution (\ref{equil}) with $\mu \neq 0$, {\it i.e.}, with a finite correlation length $\lambda_c$.

The dynamical formation of the condensate has been studied using self-similar solutions of the kinetic equation (\ref{boltz}), leading to a finite time singularity for some time $t_*$~\cite{Semikoz-95,Lacaze}. Although this singularity is naturally regularized in the NLS dynamics, it describes explicitly particle cumulation to zero wave-number ${\bf k}=0$. When the condensate begins to form ($t > t_*$), an exchange of mass between the condensate and the incoherent wave-component is necessary to reach equilibrium.
It was argued in Refs.~\cite{Pomeau-92,Dyachenko-92} that this dynamics may be described by a three-wave kinetic interaction (the fourth wave being the condensate). 
This aspect has been recently expressed more precisely by extending the kinetic equation (\ref{boltz}) to singular distributions $n_{\bf k}(t) = n_0(t) \delta^{(3)}({\bf k}) + \phi_{\bf k}(t)$~\cite{Semikoz-95,Lacaze}, so that a pair of coupled kinetic equations for the evolution of the condensate ($n_0$) and the incoherent wave-component ($\phi_{\bf k}$) has been derived~\cite{Lacaze}.
These equations describe a flux of mass from the incoherent component towards the condensate, until thermal equilibrium is reached, i.e., until the collision terms exactly vanish. This occurs for the equilibrium distribution $\phi^{eq}_k= T/k^2$, which actually corresponds to the distribution (\ref{equil}) with zero chemical potential. This allows us to legitimately assume $\mu=0$ below the transition threshold $E \le E_{{\rm tr}}$.

The number of condensed particles $n_0$ and the energy $E(\le E_{\rm tr})$ may then be calculated by setting $\mu=0$ in the equilibrium distribution~(\ref{equil}). One readily obtains $(N-n_0)/V = 
4\pi Tk_c$ and
$E/V = 4 \pi Tk_c^3/3$,
which gives 
\begin{equation} 
n_0/N = 1- E/ E_{{\rm tr}}, \label{n0}
\end{equation}
or $n_0/N=1-T/T_{\rm tr}$, where $T_{\rm tr}=3 E_{{\rm tr}}/(4 \pi V k_c^3)$.
As in standard Bose-Einstein condensation, $n_0$ vanishes at the critical temperature  
$T_{\rm tr}$, and $n_0$ becomes the total number of particles as 
$T$ tends to $0$. 
The linear behavior of $n_0$ vs. $E$ in Eq.(\ref{n0}) is consistent with the results of
numerical simulations (see Fig.~1). However note that Eq.(\ref{n0}) is derived for a 
spherically symmetric 
continuous distribution of $n_{{\bf k}}$, while in the numerics the integration 
is discretized. 
Eq.(\ref{n0}) should thus be replaced by
\begin{equation} 
{n_0 \over N} = 1- {E \over N} \ \frac{  \sum_k' 1/(k_x^2+k_y^2 + k_z^2)}{  \sum_k'  1}  
\label{n0.gen}
\end{equation}
where $\sum_k'$ denotes a discrete sum for $-k_c\leq k_x,k_y,k_z \leq k_c $ that excludes the origin $k_x=k_y=k_z =0$.

\begin{figure}[tc]
\begin{center}
\includegraphics[width=8cm, height=5cm]{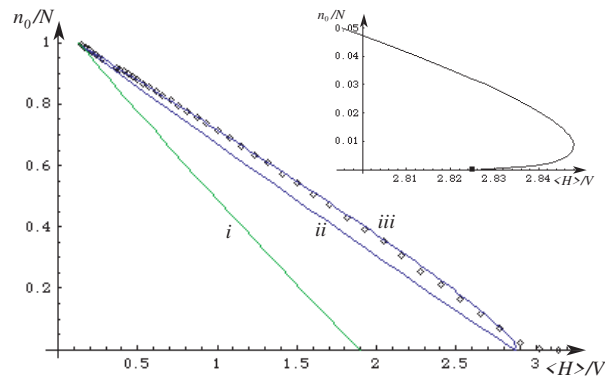}
\caption{ \label{fig1} 
Condensate fraction $n_0/N$ vs. total energy density $\left<H\right>/V$, where $\left<H\right>=E+E_0$, $E_0$ being the condensate energy [see Eq.(\ref{n0.gen2})]. 
Points ($\diamond$) refer to numerical simulations of the NLS Eq.(\ref{NLS}) with $64^3$ modes for a fixed density $N/V=1$ ($k_c= \pi$).
The straight line ({\it i}) [({\it ii})] corresponds to the continuous Eq.(\ref{n0}) [discretized Eq.(\ref{n0.gen})] approximation. Curve ({\it iii}) refers to  condensation in the presence of nonlinear interactions [from Eq.(\ref{n0.gen2})], which makes the transition to condensation subcritical, as illustrated in the inset (with $1024^3$ modes).}
\end{center}
\end{figure}

This distribution is plotted in Fig.\ref{fig1} and compared with that given by the numerical simulations of Eq.(\ref{NLS}).  In our simulations we started with a non-equilibrium distribution $\psi({\bf x},t=0)=\sum_k a_{{\bf k}} \exp(i {\bf k} \cdot {\bf x})$, where the phases of the complex amplitudes $a_{\bf k}$ are distributed randomly~\cite{Davis-prl,Berloff-02}. The numerical simulations confirm the existence of the condensation process for sufficiently low energy densities \cite{Davis-prl}. 
We performed simulations with different number of computational modes 
($8^3$, $16^3$, $32^3$, $64^3$ and $128^3$). Our numerical results reveal that once the number of modes exceeds $16^3$, it only weakly affects the 
condensate fraction $n_0/N$. This means that the system has reached some 
thermodynamic limit with only $16^3$ modes.
Moreover, we verified that for a given total volume $V$ and a fixed density $N/V$, the energy threshold for condensation increases, as the frequency cut-off increases, in agreement with the theory ($E_{\rm tr} \propto k_c^2$).

The linear dependence (\ref{n0.gen}) between $n_0$ and $E$ gives a poor approximation of the numerical results, mainly because the condensate fraction has been calculated by taking into account only the linear kinetic contribution $E$ to the total energy of the wave $H$.
To include the nonlinear (interaction) contribution, we adapt the Bogoliubov's 
theory of a weakly interacting Bose gas~\cite{bog} to the classical wave 
problem considered here.
We start from the total energy $H$ of the nonlinear wave
\begin{displaymath}
H = \sum_k k^2 a^*_{\bf k} a_{\bf k} +
\frac{1}{2V}
\hspace{-0.5cm}\sum_{k_1 , k_2 , k_3 , k_4} \hspace{-0.5cm} a^*_{{\bf
k}_1} a^*_{{\bf k}_2} a_{{\bf k}_3} a_{{\bf k}_4} \delta_{{\bf k_1}\!
+\! {\bf k_2}\! -\! {\bf k_3}\! -\!  {\bf k_4}},\label{hamiltonian}
\end{displaymath}
where $\delta_{\bf k}$ is the Kronecker delta symbol.
The Hamiltonian may be decomposed into
four terms, $H=H_0 + H_2 + H_3 + H_4$,
depending on how the zero mode, $a_0=a_{{\bf k}=0}$, and non-zero modes,
$a_{{\bf k} \neq 0}$, enter the expansion:
\begin{eqnarray}
H_0 &=&  \frac{1}{2 V}\left( |a_0|^4 + 2 |a_0|^2(N-|a_0|^2) \right) \nonumber
\\
H_2 &=&  \sum_k' \left[ \left( k^2 +  {|a_0|^2 \over V} \right)  a^*_{\bf k} a_{\bf k}  + \frac{1}{2 V}
 \left(a^2_0a^*_{\bf k}  a^*_{-{\bf k}}  + c.c\right) \right]  \nonumber
\\
 H_3 &=& \frac{1}{2 V} \sum_{k_1 , k_2 , k_3 }'
(2 a_0 a^*_{{\bf k}_1}  a^*_{{\bf k}_2} a_{{\bf k}_3} +
c.c.) \delta_{{\bf k}_1 + {\bf k}_2 - {\bf k}_3} \nonumber
\\
 H_4 &=&  \frac{1}{2V} \!\!\!
\sum_{k_1 , k_2 , k_3 , k_4}' \!\!\!\!\!\! a^*_{{\bf k}_1}  a^*_{{\bf
k}_2} a_{{\bf k}_3} a_{{\bf k}_4}  \delta_{{\bf k}_1\!+\!{\bf
k}_2\!-\!{\bf k}_3\!-\!{\bf k}_4},  \nonumber
 \label{hamiltonian2}\end{eqnarray}
where $ \sum_k'$ excludes the ${\bf k}=0$ mode.
The kinetic equation requires the Hamiltonian to be diagonal in quadratic
terms. To this end, we apply the Bogoliubov's transformation for the canonical
variables $b_{\bf k} = u_{\bf k} a_{\bf k} - v_{\bf k} a^*_{-{\bf
k}}$, with the
condition $ |u_{\bf k}|^2 - |v_{\bf k}|^2=1$ that preserves the
Poisson's bracket relation
$\{a_{\bf k},a^*_{\bf k}\}=i$ in the $b_{\bf k}$'s basis. Imposing that the
quadratic term $H_2$
is diagonal in this basis, we find
$u_{\bf k} = 1/\sqrt {1-L_{\bf k}^2}$ and $v_{\bf k} = L_{\bf k}/\sqrt {1-L_{\bf k}^2}$
with $L_{\bf k} =  [- k^2 - 2 \rho_0 + \omega_B(k)]/ \rho_0$
and $H_2={\sum_k}' \omega_B(k)\, b_{\bf k}^* b_{\bf k} $, where
$\rho_0 =  n_0/V \equiv |a_0|^2/V$, and $\omega_B(k) = \sqrt{ k^4+ 2 \rho_0  k^2}$ is the Bogoliubov's dispersion relation that takes
into account the nonlinear interaction.

Let us emphasize that, in the $b_{\bf k}$'s basis, the kinetic equations
describing the coupled evolution of the condensate $(n_0)$ and the incoherent
wave-component $(\varphi_{\bf k})$, are similar to those derived in Ref.\cite{Lacaze},
but  the linear dispersion relation $\omega(k)=k^2$ is replaced
by the Bogoliubov's expression $\omega_B(k)$. The equilibrium
distribution turns out to be $\varphi_k^{eq} = T/\omega_B(k)$, with
$\left<b_{\bf k} b_{\bf k'}^* \right> = \varphi_k^{eq} \delta({\bf
k}-{\bf k'})$,
so that the uncondensed mass (in the $b_{\bf k}$'s basis) reads
\begin{equation}
N-n_0 = \sum_k' \frac{k^2 + \rho_0}{\omega_B(k)}    \varphi_k^{eq}
= T \sum_k' \frac{k^2 + \rho_0}{\omega_B^2(k)}.
\label{n0bog}
\end{equation}
The total averaged energy $\left<H\right>$ 
has contributions from $H_0$, $H_2$ and $H_4$~\cite{foot}:
$\left<H\right> =  
E_0+  \sum_k'  {\omega_B(k)} \varphi_k^{eq}
= E_0
+  T   \sum_k'  1$, where $E_0=\frac{1}{2 V}\left[ N^2+ (N-n_0)^2 \right]$ is the energy of the condensate. 
The temperature $T$ may be substituted from this expression by using
Eq.(\ref{n0bog}), which gives a closed relation between $\left<H\right>$ and $n_0$,
\begin{equation}
\left<H\right> =  E_0
+  \frac{(N-n_0) \, \sum_k' 1}{  \sum_k' [(k^2 + \rho_0)/\omega_B^2(k)]}.
\label{n0.gen2}
\end{equation}
This expression is in quantitative agreement with the
numerical simulations of the NLS Eq.(\ref{NLS}), without any adjustable
parameter (see Fig.~1).
Expression~(\ref{n0.gen2}) remarkably reveals that the 
nonlinear interaction changes the nature of the transition to condensation,
which becomes of the first order.
This {\it subcritical} behavior 
is more pronounced when the number of modes increases. 

In summary, by considering the defocusing NLS equation as a model, we showed that a classical nonlinear wave exhibits a subcritical condensation process in 3$D$, while no transition occurs in 2$D$.
In spite of the formal reversibility of the NLS equation, the condensation process manifests itself by means of an irreversible evolution towards a homogenous plane-wave (condensate), with small-scale fluctuations superimposed (uncondensed particles), which store the information necessary for time reversal.
We formulate a thermodynamic description of the condensation process, whose properties are analogous to those of standard Bose-Einstein condensation in quantum systems.
However, caution should be exercised when drawing conclusions about condensation in real bosonic systems, because the NLS equation only describes highly occupied modes satisfying $n_{\bf k} \gg 1$, so that it cannot describe any transfer of mass between scarcely occupied high-energy modes and the condensate~\cite{Davis-01,Davis-prl,Berloff-02,Lacaze}. Such a transfer is crucial at least at the
beginning of the growth of the condensate that starts with zero mass.

To conclude we would like to mention that the equilibrium distribution (\ref{equil}) also describes the velocity spectrum distribution in 2$D$ fluid turbulence~\cite{Hasegawa}. 
An analogy between 2$D$ fluid turbulence and our results may be helpful
in the context of decaying turbulence at very high Reynolds number~\cite{Tab}. 
In this case, following Onsager's argument \cite{Onsag}, the large scale 
vortices formed by the inverse cascade would play the role of the condensate.



CC acknowledges the support of Marie--Curie grant HPMF-CT-2002-02004 and SR that of FONDECYT 1020359, Chile.


\pagebreak







\begin{thebibliography}{}
\newpage

\bibitem{Pet-85} V. I. Petviashvili and V. V. Yankov, Rev. of Plasma Phys. {\bf 14}, 5 (1985).
\bibitem{Zakharov-88} V. E. Zakharov {\it et al.,} Pis'ma Zh. Eksp. Teor. Fiz. {\bf 48}, 79 (1988) [JETP Lett. {\bf 48}, 83 (1988)]; 
S. Dyachenko {\it et al.,} Zh. Eksp. Teor. Fiz. {\bf 96}, 2026 (1989) [Sov. Phys. JETP {\bf 69}, 1144 (1989)].
\bibitem{Jordan-00} R. Jordan, B. Turkington and C. L. Zirbel, Physica D  {\bf 137}, 353 (2000); R. Jordan and C. Josserand, Phys. Rev. E  {\bf 61}, 1527 (2000).
\bibitem{Rumpf-01} B. Rumpf and A. C. Newell, Phys. Rev. Lett.  {\bf 87}, 054102 (2001); Physica D  {\bf 184}, 162 (2003).
\bibitem{Davis-01} Yu. Kagan and B. V. Svistunov, Phys. Rev. Lett.  {\bf 79}, 3331 (1997); M. J. Davis, R. J. Ballagh and K. Burnett, J. Phys. B {\bf 34}, 4487 (2001).
\bibitem{Davis-prl} M. J. Davis, S. A. Morgan and K. Burnett, Phys. Rev. Lett. {\bf 87}, 160402 (2001); Phys. Rev. A {\bf 66}, 053618 (2002).
\bibitem{Berloff-02} N. G. Berloff and B. V. Svistunov, Phys. Rev. A {\bf 66}, 013603 (2002).
\bibitem{Pomeau-92} Y. Pomeau, Physica D  {\bf 61}, 227 (1992).
\bibitem{Dyachenko-92} S. Dyachenko, A. C. Newell, A. Pushkarev and V. E. Zakharov, Physica D  {\bf 57}, 96 (1992).
\bibitem{ZakhBook} V. E. Zakharov, V. S. L'vov and G. Falkovich,
{\it Kolmogorov Spectra of Turbulence I} (Springer, Berlin, 1992); A. C. Newell, S. Nazarenko and L. Biven, Physica D {\bf 152}, 520 (2001).
\bibitem{Chiao} R. Y. Chiao {\it et al.,} Phys. Rev. A  {\bf 69}, 063816 (2004).
\bibitem{Boyd} R. W. Boyd, {\it Nonlinear Optics} (Acad. Press, 2002).
\bibitem{zakhprl04} 
See, e.g., E. Henry, P. Alstrom and M. T. Levinsen, Europhys. Lett. {\bf 52}, 27 (2000);
A. I. Dyachenko, A. O. Korotkevich, and V. E. Zakharov, Phys. Rev. Lett. {\bf 92}, 134501 (2004).
\bibitem{Robinson-97} P. A. Robinson, Rev. Mod. Phys. {\bf 69}, 507 (1997).
\bibitem{GP} V. L. Ginzburg and L. P. Pitaevskii, Sov. Phys. JETP {\bf 7}, 858
(1958); L. P. Pitaevskii, Sov. Phys. JETP {\bf 13}, 451 (1961); E. P. Gross J. Math. Phys. {\bf 4}, 195 (1963).
\bibitem{Semikoz-95} B. V. Svistunov, J. Mosc. Phys. Soc. {\bf 1}, 373 (1991); D. V. Semikoz and I. I. Tkachev, Phys. Rev. Lett.  {\bf 74}, 3093 (1995); Phys. Rev. D {\bf 55}, 489 (1997).
\bibitem{Lacaze} R. Lacaze {\it et al.,} Physica D {\bf 152}, 779 (2001); C. Connaughton and Y. Pomeau, C. R. Physique {\bf 5}, 91 (2004).
\bibitem{bog} N. N. Bogoliubov, Journal of Physics {\bf 11}, 23 (1947).
\bibitem{foot} The $H_4$ term 
is not of higher order than $H_0$ or $H_2$. After a rapid inspection, one sees that particular terms where all four $k_\alpha$'s ($\alpha=1,2,3,4$) are the same, or those where:
$k_1 = k_3$, $k_2 = k_4$ (and $3\leftrightarrow 4$) contribute up to a first order. Other terms introduce correlations which are treated through the random phase approximation. The final sum in $H_4$ reads 
$ H_4 =\frac{1}{2V} \left( \sum_{\alpha}'
|a_\alpha |^4 + 2  \left(\sum_{\alpha}'
|a_\alpha |^2 \right)
\left(\sum_{\beta}' |a_\beta |^2\right)   \right). $ 
However, the $\sum_{\alpha}' |a_\alpha |^4$ is definitively negligible.
\bibitem{Hasegawa} A. Hasegawa, Adv. in Phys. {\bf 34}, 1 (1985).
\bibitem{Tab} G. F. Carnevale {\it et al.,} Phys. Rev. Lett. {\bf 66}, 2735 (1991);
O. Cardoso, D. Marteau and P. Tabeling, Phys. Rev. E {\bf 49}, 454 (1994).
\bibitem{Onsag} L. Onsager, Nuovo. Cim. Suppl. {\bf 6}, 279 (1949).

\end{thebibliography}
\end{document}